\input{aipcheck}

\documentclass[
    ,final            
  ]
  {aipproc}

\layoutstyle{6x9}

\begin{document}

\title{The SN 1987A Link to Gamma-Ray Bursts}


\author{John Middleditch}{
  address={Address: MS B265, Los Alamos National Laboratory, Los Alamos, NM 87545}
}



\begin{abstract}
Early measurements of SN 1987A can be interpreted in light of the 
beam/jet (BJ) which had to hit polar ejecta (PE) to produce the 
``Mystery Spot'' (MS), some 22 light-days distant.
It takes an extra {\it eight} days for the
SN flash to hit the MS, and early measurements confirm 
2$\times$10$^{39}$ ergs/s in the optical for a day at day 8, before 
dropping off by day 8.5.  A linear ramp in luminosity starting near 
day 10 indicates particles from the BJ hitting the PE, with the fastest 
particles traveling at 0.8 c, and an upper limit for the optical 
luminosity of the MS of 5$\times$10$^{40}$ ergs/s, about 23\% of the 
total of 2.1$\times$10$^{41}$ ergs/s at day 20.  The many details of 
SN 1987A strongly suggest that it resulted from a merger of two stellar 
cores of a common envelope (CE) binary, i.e. a ``double degenerate''
(DD)-initiated SN.  Without having to blast through the CE of Sk 
-69$^{\circ}$ 202, it is likely that the BJ would have caused a
full, long-soft gamma-ray burst ($\ell$GRB) upon hitting the PE, 
thus DD is a mechanism which can produce $\ell$GRBs.  A 0.5$^{\circ}$ 
offset, the typical collimation for a GRB, over the 22 light-days 
from SN 1987A to its PE, produces $\sim$100 s of delay, {\it matching}
the observed delay of the non-prompt parts of $\ell$GRBs.  Because
DD must be the overwhelmingly dominant SN mechanism in elliptical 
galaxies (EGs), where only short, hard GRBs (sGRBs) have been observed, 
DD without CE or PE must also produce sGRBs, and thus the initial photon 
spectrum of 99\% of {\it all} GRBs is {\it known}, and neutron star 
(NS)-NS mergers may not make GRBs as we know them.  Millisecond pulsars 
(MSPs) in the non-core-collapsed globular clusters are also 99\% 
DD-formed from white dwarf (WD)-WD merger, consistent with their 2.10 
ms minimum spin period, the 2.14 ms signal seen from SN 1987A, and sGRBs
offset from EGs.  The 
many details of Ia's strongly suggest that these are also 
DD.  This is a concern for systematics in Ia Cosmology, because Type Ia 
SNe will appear to be Ic's when viewed from their DD merger poles, given 
sufficient matter above that lost to core-collapse (otherwise it would 
just beg the question of what {\it else} they could possibly be, and 
other mechanisms, such as ``delayed detonation'' or ``gravitationally 
confined detonation,'' don't produce the inverse relation between 
polarization and luminosity).  As a DD-initiated SN, SN 1987A appears to 
be the Rosetta Stone for 99\% of SNe, GRBs and MSPs, including all 
recent nearby SNe except SN 1986J, and the more distant SN 2006gy.
There is no need to invent exotica to account for GRBs.
 


\end{abstract}

\maketitle


\subsubsection{Introduction}

Gamma-ray bursts (GRBs) are the most luminous objects in the Universe,
yet we still know very little about them (see \cite{Mz06} and references 
therein), although some have been found to be associated with
SNe but others, mostly
those lasting only a fraction of a second, with slightly harder
spectra, produce only ``afterglows,'' sometimes
extending down to radio wavelengths.  A large number of models
have been put forth to explain GRBs, including NS-NS
mergers for some, and many flatout {\it inventions}, such as
``collapsars,'' ``hypernovae,'' and ``supranovae,'' for others.
This work offers  a simple explanation for 99\% of SNe, GRBs, {\bf and} 
MSPs, in the context of a phenomenon that Nature 
has {\it already} provided, namely SN 1987A (87A).

\subsubsection{The SN 1987A Bipolarity and ``Mystery Spot''}

The explosion of 87A is clearly bipolar \cite{NASA06,Wang02}.
A ``polar blowout feature''
(PBF -- the prime suspect for the r-process) approaches at about 
45$^{\circ}$ off our line of sight.
It partially obscures an equatorial bulge/ball (EB), behind which 
a part of the opposite, receding PBF is visible.  The 87A PBFs 
and EB are approximately equally bright, in contrast to what
polarization observations imply for Type Ia SNe (see below).

A binary merger scenario of two electron degenerate stellar cores 
(double degenerate, or DD) has been proposed for 87A,
and the triple ring struture has recently been consistently calculated
\cite{MP07}.
The many other details of 
87A, including the mixing, the blue supergiant progenitor,
and the 2.14 ms pulsed optical signature, strongly support this
hypothesis.  Prior to and during this time 
{\it no} measurement of {\it any} SN other than SN 1986J,\footnote{This
SN occurred in the edge-on spiral galaxy, NGC 0891, and exceeds
the luminosity of the Crab nebula at 15 GHz by a factor of 200, and thus
is thought to have occurred because of the iron photodissociation
catastrophe (Fe PdC) mechanism, thereby producing a {\it strongly}
magnetized NS (the origin of magnetic fields in NSs is still poorly
understood, though it is believed that thermonuclear [TN] combustion 
in the massive progenitor to an Fe core is related).  As a corollary, 
we note that models of SNe to date have not taken DD into account, 
and {\it certainly} have not been calibrated to an Fe PdC SN, such as 
1986J.  The inner layers of all Fe PdC SNe, possibly {\it many} 
M$_{\odot}$ of Si, C, and O, have not been diluted with H and/or He by 
DD, and thus may ignite upon core-collapse and burn efficiently to 
$^{56}$Ni, making a strongly 
magnetized remnant a possiblity even for SN 2006gy \cite{Sm07}, which 
may have produced $\sim$20 M$_{\odot}$ of $^{56}$Ni.} or of {\it any} 
GRB other than SGRs,
has been inconsistent with this geometry.

The first clear evidence for DD-formed MSPs
coincidentally came in the birth year of 87A, with the discovery 
of the 3 ms pulsar, B1821-24 \cite{Lyne}, in the non-core-collapsed 
(nCCd) globular cluster (GC) M28.
This discovery was followed by many, many more in the nCCd GCs, such as
47 Tuc, over the last 20 years, and attributing these to recycling through 
X-ray binaries has never really worked \cite{CMR}, by a few 
orders of magnitude.\footnote{Recycled pulsars weighing 1.7 M$_\odot$, 
in the CCd GC, Ter 5 (Scott Ransom 2006), have removed high accretion 
rate from contention as a alternative mechanism to produce the MSPs in 
the nCCd GCs.}~ Other works \cite{M00b,M04} have explored DD
more recently.

The most remarkable
feature\footnote{Not counting, for the moment, the 2.14 ms pulsed
optical remnant, which also revealed a $\sim$1,000 s precession 
\cite{M00a,M00b}.  Since a prototypical, dim, thermal neutron star
remnant (DTN) has been discovered in Cas A \cite{Ta99}, representing what
PSR 1987A will look like after another 300 years, and other pulsars have since
been observed to precess \cite{St00}, this candidate is no longer
controversial.}
of 87A was the ``mystery spot'' (MS -- Fig.~1), with a thermal
energy of 10$^{49}$ ergs, even 50 days {\it after} the CC event 
and separated from the SN photosphere ``proper'' by some 0.06 arc s 
{\it along the axis of its DD merger} toward the Earth, some 45$^{\circ}$ off 
our line of sight, with about 3\% of this energy eventually radiated
in the optical band.  The geometry is such that it takes light only about 
{\it eight} extra days to hit it and continue on to be observed from the 
Earth (Fig.~1).

The approaching polar beam/jet produced by 87A, which blasted
through the CE of Sk -69$^{\circ}$ 202 and 
collided with what is thought to be PE some 22 light-days distant from the 
SN along this axis, may be {\it generic} to the DD process.
Through its interaction with the overlaying CE and/or PE it 
produces the wide variation in GRB/X-ray flash properties 
observed from DD SNe of sufficiently low inclination to the line of sight.
Thus the flavors of the 99\% of GRBs due to DD depend on CE and/or PE
mass (Fig.~2, left).  The restricted T$_{90}$ range of the new iGRB
class is consistent with DD events within RSGs, and the early
polarization of SN 1993J \cite{THW93}, at 1.6\% and 1.0\%, being
twice that of 87A (0.9\% and 0.4\% \cite{Ba88}).
Since (per the Abstract) DD also produces sGRBs, it is 
not clear if {\it any} of these result from NS-NS mergers,
bad news for the Earth-based gravitational observatories.  Without
circum-merger material, this process completes in a few milliseconds,
shorter than the shortest sGRBS, out of range of Fig.~2, left.

Following the initial flash, very early data on 87A 
(Fig.~2, right) shows the breakout of the hotter, central
part of the BJ by day 3, then cooling (or losing the ability
to do so) until day 8, when the flash scatters in the PE
(etc., per the Abstract).  

\begin{figure}
  \includegraphics[height=.3\textheight]{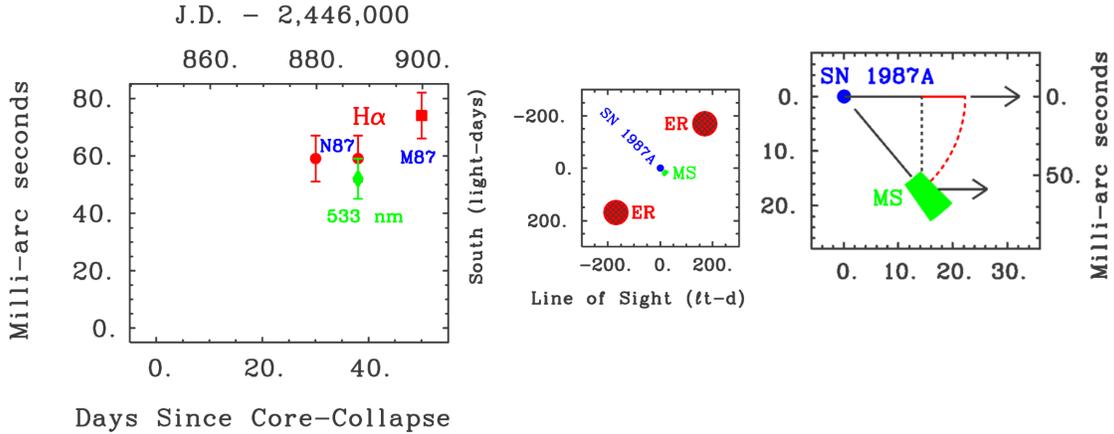}
\caption{(Left) The distance of the MS from 87A at 
H${\alpha}$ and 533 nm vs time, from \cite{M87,N87}
(M87/N87).  (Center) The MS geometry
and cross-sections of the equatorial ring (ER).  (Right) The
lines of sight from 87A to the Earth directly/through the 
MS (which takes an extra 8 days).  }
\end{figure}

\begin{figure}
  \includegraphics[height=.3\textheight]{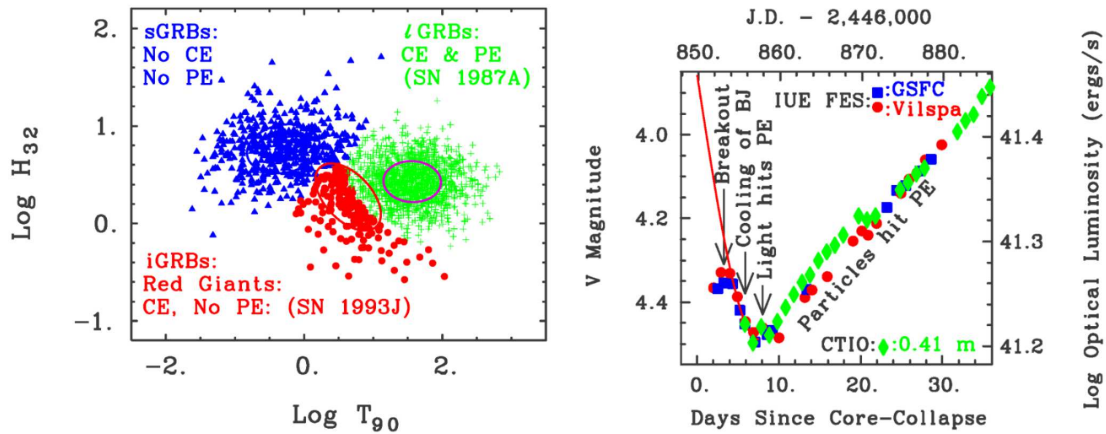}
\caption{(Left) After \cite{H06}
the GRBs from the BATSE
Catalogue \citep{Me01} scattered in duration (T$_{90}$)-hardness
(H$_{32}$) space.  (Right) After \cite{W87} (Vilspa),
with data from Sonneborne \& Kirshner added (GSFC), the very early light 
curve of 87A from the IUE FES, together with data from the 
CTIO 16-inch telescope \citep{HS90} (the centers of the two V bands are 
5,100 \AA~\& 5,500 \AA~respectively).
}
\end{figure}

\begin{figure}
  \includegraphics[height=.3\textheight]{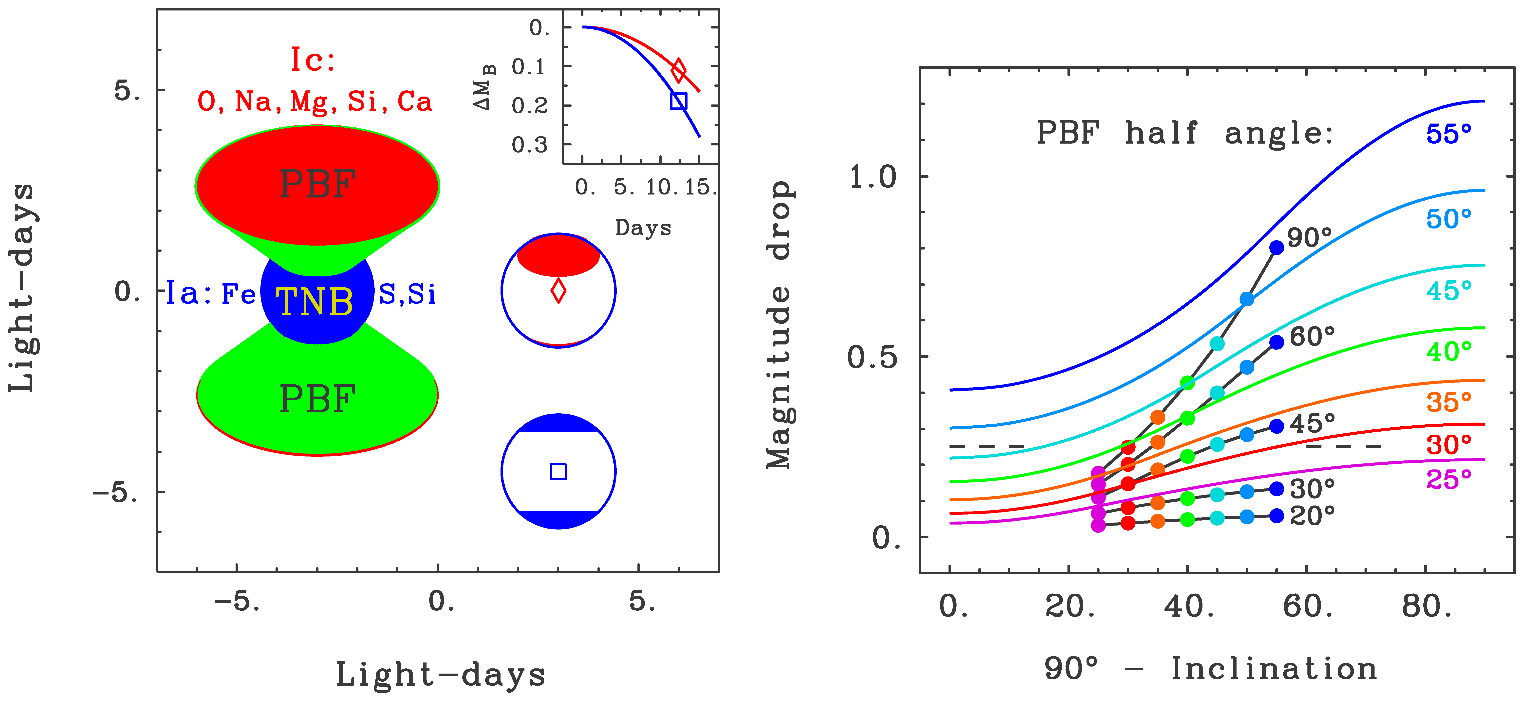}
\caption{(Left) The geometry for type Ia/c SNe, as viewed 30$^{\circ}$
off the merger equator, with PBFs sketched as cones of half-angle 
45$^{\circ}$.  The two circles at the right/lower right show
the maximum PBF footprints on the much brighter TNBs, which can be
exposed as the PBFs quickly depart, for merger co-inclinations
(co-i's) of 0$^{\circ}$ (square) and 30$^{\circ}$ (diamond).  These
effects are compounded with a parabolic $\Delta$M$_{B{_{15}}}$
of 0.5 m in 15 days (inset).  (Right) The maximum drop in magnitude
from exposure of PBF footprints on the TNB as a function of
co-i for PBF half angles of 25$^{\circ}$-55$^{\circ}$
in 5$^{\circ}$ steps (curves without dots), and the excess
drops over 0$^{\circ}$ for co-i's of 20$^{\circ}$, 30$^{\circ}$, 45$^{\circ}$,
60$^{\circ}$, and 90$^{\circ}$ (curves with dots), plotted at
the abscissa values equal to their PBF half angles.
}
\end{figure}

\subsubsection{DD in Type Ia/c SNe}

{\it Every} observation of Type Ia SNe ever made is consistent with the 
bipolar explosion geometry of 87A, thus it seems reasonable to 
suggest that these too are DD-initiated SNe, which still produce 
TN yield, but also leave weakly magnetized MSPs, rather than single 
degenerate, disrupting TN blasts.  The list of good reasons 
supporting the DD hypothesis for Ia's is long:
(1 \& 2) no SN-ejected or companion wind-advected H/He, (3) ubiquitous 
high velocity features (HVFs), (4 \& 5) SiII/continuum polarization (CP) 
both $\propto$1/luminosity (IPL), (6) no radio Ia SNe, (7) {\it four} 
Ia's/(26 years) in the merging spiral/elliptical galaxies comprising 
NGC 1316, (8) $>$1.2 M$_{\odot}$ of $^{56}$Ni in SN 2003fg, (9) 
cataclysmic variables are explosive, \& (10) DD SNe are needed to 
account for the abundance of Zinc -- see the references in \cite{M06}.  

Thus Ia's appear to be similar to the bipolar explosion 
of 87A, except that their TN/EB (TNB) dominates their luminosity 
over the PBFs, hence the very low, but still IPL CP ($\sim$0.0-0.5\%).  
The primary effect of Ia PBFs is likely to be their shading of the TNBs.
These PBFs have higher velocities than in 87A because of lighter
overlayers, assuming that the DD mechanism does not differ underneath, 
as seems reasonable.  

Ia's in old populations are almost exclusively due to CO-CO 
WD merger.  In younger populations they can also be merging CE
binary WR stars \cite{DeM}.  Occam's Razor and the extreme
bipolarity of Ia's implied by their HVFs and IPL
polarizations would suggest that Ia's observed from the DD merger 
poles are Type Ic SNe, given enough overlayer to shroud the TN ashes 
of Si, S, \& Fe (otherwise per the Abstract).
Thus Ia/c's form a continuous class of SNe, varying only by overlayer 
mass and observer co-i.

This is a concern for Ia Cosmology, as small levels of Ibc
contamination can spuriously produce a high fraction of the 
$\Omega_{\Lambda}$ = 0.7 \cite{Ho05}.  Figure 3 shows how
a fraction of the bright TNB can be exposed by the 
departing (and relatively dark) PBFs during the interval when 
$\Delta$M$_{15}$ is being measured, producing a SN which appears to 
be too faint for the redshift of its host galaxy, due to inadequate 
correction.  The effect for a 30$^{\circ}$ co-i and PBFs 
with 1/2 angles of 45$^{\circ}$, as drawn in Fig.~3 (left), accounts 
for {\it half} of $\Omega_{\Lambda}$ = 0.7,
{\it even though this difference
is scarcely apparent to the eye} 
(the dashed line segments at
0.25 mag in Fig.~3, right, represent the {\it whole} effect).  
More realistic TNBs which begin as toroids may produce a big
effect even for low co-i's.

Have overdiligent attempts to select a local sample of Ia's 
``uncontaminated'' by Ic's, excluded events in error by a tenth
to a whole magnitude, which are {\it not} routinely excluded from
distant samples?  There appears to be a gap between the Ia's
underluminous by 1-2 whole magnitudes \cite{Be05}, which
are easily excluded by the TiII $\lambda \lambda$ 4,000-4,500
\AA~shelf, and the others in the local sample.  Are
there Ia's in this gap which may not be so easily exluded 
from the distant samples?~ Time will tell.

In conclusion, 87A is the Rosetta Stone for 99\% of SNe,
GRBs, and MSPs, and NS-NS mergers may not make GRBs as we know 
them.  I thank the Observatories of the Carnegie Institution of
Washington for support (see \cite{M06} for
further acknowledgments).

\bibliographystyle{aipprocl} 


\IfFileExists{\jobname.bbl}{}
 {\typeout{}
  \typeout{******************************************}
  \typeout{** Please run "bibtex \jobname" to optain}
  \typeout{** the bibliography and then re-run LaTeX}
  \typeout{** twice to fix the references!}
  \typeout{******************************************}
  \typeout{}
 }

\end{document}